# Autoencoder for Interconnect's Bandwidth Relaxation in Large Scale MIMO-OFDM Processing

M. Ahmed Ouameur and D. Massicotte

*Abstract—* **Deep learning is playing an instrumental role in the design of the next generation of communication systems. In this letter, we address the massive MIMO interconnect's bandwidth constraint relaxation using autoencoders. The autoencoder is trained to learn the received signal structure so that a low dimension latent variable is transferred as opposed to the original high dimension signal. For an efficient implementation, the approach suggests to separately deploy the autoencoder components, namely the encoder and the decoder, in massive MIMO radio head and central processing units respectively. The simulation results show that one can relax the interconnect's bandwidth by a factor of up to 8 with non-substantial performance degradation of the centralized processing in comparison with the decentralized processing running with four clusters.**

## I. INTRODUCTION

Now that massive MIMO is a reality and has already made its way to 5G, five promising research directions are discussed in [1]. Therein machine learning (ML) has been identified as an indispensable tool to enable intelligent massive MIMO communication. This vision is inherently supported by the authors in [2] where the main question is no longer whether deep learning (DL) will be integrated into next-generation radio access networks but rather how and when. Because of its advantages in terms of high spectral efficiency, increased reliability, and power efficiency, massive MIMO has been the subject of a large number of research activities [3]. It has also been advocated in [1] that going extremely massive, with hundreds or thousands of antennas, is one of the promising directions to provide order-of-magnitude higher area throughput in wireless networks.

Recent contributions seem to promote the potentials of using DL for communication system design [4]-[7]. Even if most signal processing algorithms have solid and well-established roots in statistics and information theory for tractable mathematical models, it remains that a practical system has many impairments and non-linearities, which can be roughly captured by such models [8]. For this reason, a DL-based communications system, which is tailored for a specific hardware configuration and channel, might be able to better optimize in the presence of such impairments [9].

On the other hand, as the number of antennas increases towards building an extremely massive MIMO system, favorable propagation channel conditions are expected [3], [10]. As such, the users' channels are mutually orthogonal which makes linear processing based on maximum ratio combining (MRC), zero-forcing (ZF) detection or minimum mean squared error (MMSE) detection, a suitable and optimal choice [10]. Many works on linear and low complexity processing are proposed in [11], [12]. Nevertheless, all these centralized processing techniques still impose stringent constraints on the interconnects' bandwidth between the massive MIMO radio heads (RHs) and the central processing unit (CPU). Distributed, or decentralized, massive MIMO processing has been introduced to overcome such limitation [13]-[15] and [17]. Unfortunately, the decentralized processing computational complexity, and hence the energy efficiency, is also of concern. We, therefore, attempt to return back to the original centralized processing problem and figure out a means to efficiently reduce the data transfer rate as close to that proposed by decentralized processing while having similar performance. In doing so, one would expect to enjoy centralized processing advantages such as low computational complexity and latency.

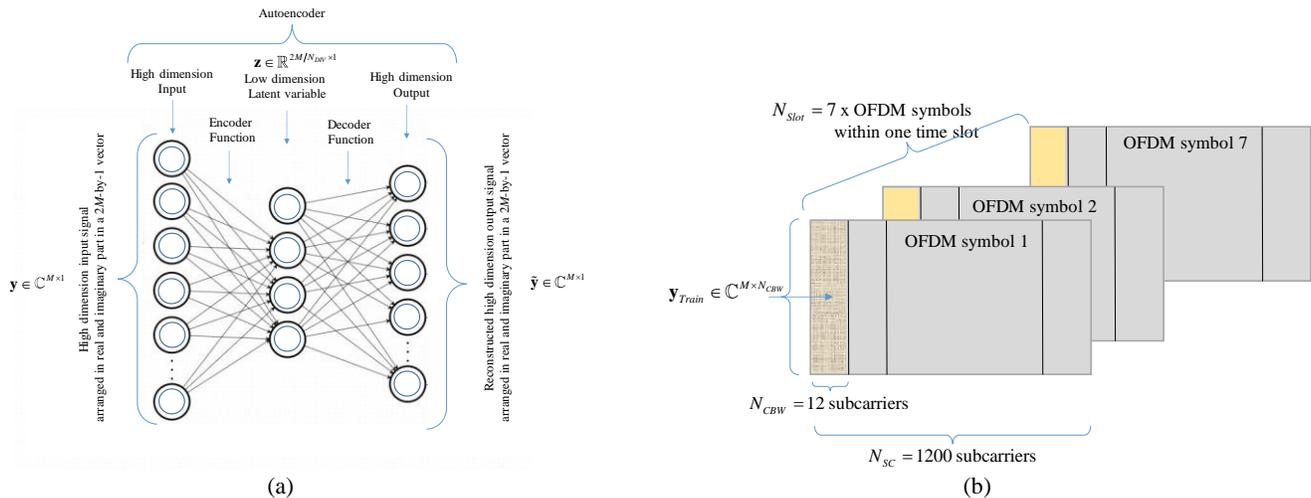

(a)                                        (b)

Figure 1. Proposed autoencoder concept (a) and the received OFDM data frame (time slot) after synchronization and FFT operation per antenna element (b).

While moving towards an intelligent massive MIMO system [1], our approach is rather to deploy the autoencoder's encoder and decoder functions at the massive MIMO RHs and the CPU respectively. Based on the autoencoder concept, one can think of training an autoencoder with the raw high dimension received samples prior to massive MIMO processing. During the processing phase, only the low dimension latent variable is transferred between the massive MIMO RH and the CPU (see figure 1.a). During the uplink detection operation, for instance, this is achieved by deploying the encoder at the massive MIMO RH interface while the decoder is implemented at the CPU interface to reconstruct the original signal.

The main contributions of this letter are:

- Introduce the autoencoder concept (c.f. Figure 1.a) as a means to reduce the interconnect's bandwidth in centralized massive MIMO processing wherein only a low dimension latent variable is exchanged.
- We evaluate the uplink detection performance of the centralized processing with autoencoder based interconnection bandwidth reduction versus the decentralized processing with full bandwidth.

The paper is organized as follows: Section II presents the uplink signal model. Section III discusses the autoencoder concept to reduce the constraint on the interconnect's bandwidth. The impact of signal reconstruction loss on the uplink massive MIMO detection's performance is evaluated in Section IV considering the iterative low complexity centralized and decentralized processing. Finally, the conclusions are drawn and some future research directions are outlined in Section V.

## II. SIGNAL MODEL, CENTRALIZED AND DECENTRALIZED PROCESSING

### A. Signal model

We consider an uplink transmission where $K$ single antenna users are communicating with a BS equipped with $M$ antennas (where $M \gg K$) in TDD duplex mode using the OFDM modulation scheme (c.f Figure 1.b). For the sake of notation simplicity, we consider a baseband equivalent channel and expressions per subcarrier where the subcarrier index is suppressed. The data signal of the $k^{th}$ user is denoted by $s_k \in \mathbb{C}$ and is normalized to unit power. The vector $\mathbf{h}_k \in \mathbb{C}^{M \times 1}$ represents the corresponding channel which is modeled, for simulation purposes, as a flat Rayleigh fading channel vector whose entries are assumed to be independent and identically distributed (i.i.d) with zero mean and unit variance. We model the received signal at the BS as

$$\mathbf{y} = \mathbf{Hs} + \mathbf{n} \qquad (1)$$

where $\mathbf{y} \in \mathbb{C}^{M \times 1}$, $\mathbf{H} = \begin{bmatrix} \mathbf{h}_1 & \mathbf{h}_2 & \cdots & \mathbf{h}_K \end{bmatrix}$ is the channel matrix and $\mathbf{s} = \begin{bmatrix} s_1 & s_2 & \cdots & s_K \end{bmatrix}^T$. $\mathbf{n} \in \mathbb{C}^{M \times 1}$ represents the additive receiver noise vector whose entries have a zero mean and a variance equal to $\sigma^2$.

The zero-forcing (ZF) detection technique applies $\mathbf{W} = \left( \mathbf{H}^H \mathbf{H} \right)^{-1} \mathbf{H}^H = \begin{bmatrix} \mathbf{w}_1, & ..., & \mathbf{w}_K \end{bmatrix} \in \mathbb{C}^{M \times K}$ to the received signal $\mathbf{y}$ to estimate the users' transmitted signal $\mathbf{s}$ as

$$\hat{\mathbf{s}} = \mathbf{Wy} = \left( \mathbf{H}^H \mathbf{H} \right)^{-1} \mathbf{H}^H \mathbf{y} = \left( \mathbf{H}^H \mathbf{H} \right)^{-1} \mathbf{y}_{MF} = \mathbf{A}^{-1} \mathbf{y}_{MF} \quad (2)$$

where $\mathbf{y}_{MF} \triangleq \mathbf{H}^H \mathbf{y}$. Notice that the maximum ratio combining

TABLE 1. AUTOENCODER PARAMETERS

| Properties | |
|---|---|
| **Item** | **Value** |
| Latent Variable size | $(2M)/N_{DIV}$, $N_{DIV} = \{2,4,8,\cdots\}$ |
| Encoder transfer function | Log sigmoid |
| Decoder transfer function | Log sigmoid |
| Encoder weights | $(2M)/N_{DIV} \times N_{CBW}$ |
| Decoder weights | $N_{CBW} \times (2M)/N_{DIV}$ |
| **Training parameters** | |
| **Item** | **Value** |
| Loss function | MSE |
| Algorithm | Scaled conjugate gradient |
| Max. epochs | 10000 |
| L2 weight regularization | 0.001 |
| Sparsity regularization | 1 |
| Sparsity proportion | 0.05 |

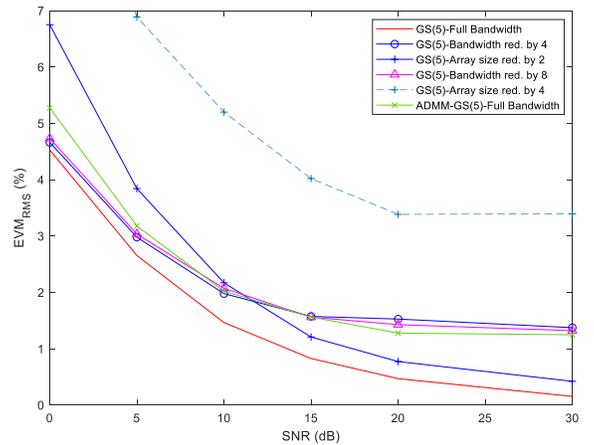

Figure 2. Centralized GS based detection [12] performance at varying interconnect's bandwidth reduction factors vs GS-ADMM [17] at full bandwidth. The performance at full bandwidth assumes the use of all received signal samples without passing them through the autoencoder.

(MRC) technique considers $\mathbf{A}^{-1} \cong \left( diag\left( \mathbf{H}^H \mathbf{H} \right) \right)^{-1}$. Where $diag\left( \bullet \right)$ operator extracts the diagonal part of its argument.

### B. Centralized and decentralized processing

To evaluate the effect of signal reconstruction loss, two massive MIMO detection approaches are considered. We resort to the Gauss-Seidel (GS) method [12] for centralized processing and to the computationally efficient ADMM-GS method [17] for decentralized processing. Due to the limited space, we refer the reader to [17] and the references therein. In related works, the authors in [13] have proposed an alternating direction method of multipliers (ADMM) based processing where a base station BS is divided into a cluster of small independent groups of RHs with fewer antenna each. By exploiting the ADMM framework [14], novel decentralized processing is suggested. Other techniques do also rely on the iterative exchange of consensus information which limits the achievable rate due to the inherently higher latency [15]. To mitigate such latency, the approach in [15] is to avoid sharing the consensus information among the clusters by proposing a decentralized feedforward architecture. Unfortunately, the decentralized processing computational complexity is still of a concern. We, therefore, revert back to the original centralized processing problem and figure out a means to efficiently reduce the data transfer rate as close to that proposed by decentralized processing while having similar performance.

### III. AUTOENCODER TRAINING AND DEPLOYMENT

At the first look, this problem may be perceived as a data compression problem. However, unlike compression techniques used for channel state information (CSI) or feedback information, which run at typically moderate 3:1 compression rate [16], the interconnect bandwidth requirements are very stringent. They require an order of magnitude reduction factor of as high as the number of clusters used in the decentralized processing counterpart. Therefore, if the array is subdivided into eight clusters one would expect the interconnect bandwidth requirement to be relaxed by a factor proportional to eight which is beyond the capabilities of the current compression techniques. In addition, CSI may exhibit some properties, such as sparsity, that can be exploited [16 and references 14-16 therein]. Unfortunately, the raw received signal samples do not seem to trivially embedded key properties which can be exploited to devise efficient compression techniques.

### A. Autoencoder training and deployment

Figure 1.a depicts an autoencoder which operates on a $2M \times 1$ real-valued signal vector comprising of the concatenated real and imaginary parts of the original signal $\mathbf{y} \in \mathbb{C}^{M \times 1}$. The reconstructed output signal $\tilde{\mathbf{y}} \in \mathbb{C}^{M \times 1}$ is of the same dimension. We assume that the massive MIMO RH is equipped with an autoencoder coprocessor (e.g. an embedded GPU). As such, during the training phase, the autoencoder coprocessor is trained to infer the encoder and the decoder functions. During the detection phase, the encoder and the decoder functions are distributed in such a way that the encoder and the decoder functions are deployed in the massive MIMO RH and the CPU respectively. If the autoencoder is configured with a low latent variable dimension, the constraint on the interconnect's bandwidth between the massive MIMO RH and CPU is greatly relaxed if the low dimension latent variable is transferred.

### B. OFDM data frame

Figure 1.b shows a typical OFDM data frame used during simulations. This is a typical 20 MHz LTE-like time slot with 7 OFDM symbols. Each OFDM symbol consists of $N_{SC} = 1200$ subcarriers per antenna. We assume that $N_{CBW} = 12$ subcarriers would span a typical channel coherence bandwidth while $N_{Slot} = 7$ OFDM symbols would span a typical channel coherence time. We define $\mathbf{y}_{Train} \in \mathbb{C}^{M \times N_{CBW}}$ as a training block. During the training phase, the autoencoder is fed with one training block $\mathbf{y}_{Train}$ of the first OFDM symbol at the beginning of each coherence block. A coherence block contains $2M \times N_{CBW} \times N_{Slot}$ worth of real-valued samples (subcarriers).

### C. A note on the effective interconnection's bandwidth reduction

Note that the number of real-valued samples to transfer within one coherence block is $2M \times N_{CBW} \times N_{Slot}$ (the factor 2 accounts for the real and the imaginary parts of the complex-valued sample/subcarrier). Using the autoencoder with $N_{DIV}$ interconnection's bandwidth reduction factor (see table 1), only the low dimension latent variable of $2M \times N_{CBW} \times N_{Slot} / N_{DIV}$ real-valued samples is transferred. Accounting for the transfer overhead of $N_{CBW} \times 2M / N_{DIV}$ the decoder's weights (see table 1), the total number of transferred real-valued samples is $2M \times N_{CBW} \times N_{Slot} / N_{DIV} + N_{CBW} \times 2M / N_{DIV}$. For the interconnection's bandwidth reduction factor $N_{DIV} = 8$, $M = 512$ antennas, $N_{CBW} = 12$ subcarriers, and $N_{Slot} = 7$ OFDM-symbols, the effective reduction factor is 7.466.

### IV. PERFORMANCE RESULTS AND ANALYSIS

This section discusses the effect of the autoencoder's reconstructed signal loss on the centralized massive MIMO detection performance. The autoencoder is trained based on the parameters in Table 1 where only one training block per coherence block is used. Therefore, only the first $\mathbf{y}_{Train} \in \mathbb{C}^{M \times N_{CBW}}$ in each OFDM data frame is used. The detection process will operate over all OFDM symbols (including the training block) so that no loss in spectral efficiency is expected. Unless otherwise stated, we consider a massive MIMO RH with $M = 512$ antennas serving $K = 40$ single-antenna users. The number of clusters (groups) in the decentralized processing is set to 4 (to keep the number of antennas per cluster higher than the number of users). The training is performed at 10 dB SNR with i.i.d channel realizations being different at subsequent subcarriers.

GS based detector's performance is evaluated with varying latent variable size. We set the latent variable size as $2M/N_{DIV}$ per subcarrier where $N_{DIV} = \{2, 4, 8, \cdots\}$ represents the interconnect's bandwidth relaxation/reduction (red.) by a factor of 2, 4 and 8. Each technique is implemented with 5 iterations (typically 3 to 5 iterations are usually used for low complexity implementations). For a fair comparison, we use a baseline configuration (without autoencoding) where the array size is scaled as $M/N_{DIV}$ (i.e. array size reduction) to align the interconnect's bandwidth on the same constraint. Figure 2 shows the error vector magnitude (EVM) as a function of the SNR for the GS based centralized detection method. Interestingly, three key remarks can be made:

(i) The autoencoder seems to have little impact on the centralized processing performance at low SNR region.

(ii) It represents a good alternative for relaxing the interconnect's bandwidth as opposed to reducing the antenna array size. It is therefore far more efficient reducing the interconnect's bandwidth by 8 using autoencoder than reducing the array size by 4.

(iii) The centralized processing with up to 8x bandwidth reduction shows similar performance compared to the decentralized processing with four clusters.

Not shown in the figure, the loss in the performance is not substantial up to $16\times$ interconnect's bandwidth constraint relaxation.

## V. CONCLUSION

This letter provides some insights on the use of deep learning (autoencoder) as a means to reduce the interconnect's bandwidth constraint for centralized massive MIMO processing. The performance loss is substantially far less than resorting to reducing the array size to cope with the constraints imposed by the physical interconnects between the massive MIMO RH and the CPU. One future direction is to apply autoencoder over more adjacent coherence blocks (e.g. one full OFDM symbol) which usually experience different propagation channel conditions. It is also feasible to consider such a strategy as part of the DL based end-to-end communication system.